# Ordered-subsets Multi-diffusion Model for Sparse-view CT Reconstruction

Pengfei Yu, Bin Huang, Minghui Zhang, Weiwen Wu, *Member, IEEE*, Shaoyu Wang, Qiegen Liu, *Senior Member, IEEE*

*Abstract*—Score-based diffusion models have shown significant promise in the field of sparse-view CT reconstruction. However, the projection dataset is large and riddled with redundancy. Consequently, applying the diffusion model to unprocessed data results in lower learning effectiveness and higher learning difficulty, frequently leading to reconstructed images that lack fine details. To address these issues, we propose the ordered-subsets multi-diffusion model (OSMM) for sparse-view CT reconstruction. The OSMM innovatively divides the CT projection data into equal subsets and employs multi-subsets diffusion model (MSDM) to learn from each subset independently. This targeted learning approach reduces complexity and enhances the reconstruction of fine details. Furthermore, the integration of one-whole diffusion model (OWDM) with complete sinogram data acts as a global information constraint, which can reduce the possibility of generating erroneous or inconsistent sinogram information. Moreover, the OSMM's unsupervised learning framework provides strong robustness and generalizability, adapting seamlessly to varying sparsity levels of CT sinograms. This ensures consistent and reliable performance across different clinical scenarios. Experimental results demonstrate that OSMM outperforms traditional diffusion models in terms of image quality and noise resilience, offering a powerful and versatile solution for advanced CT imaging in sparse-view scenarios.

## I. INTRODUCTION

COMPUTED Tomography (CT) has gained widespread use in medical diagnosis due to its ability to provide practical and precise diagnostic outcomes [1]. As image quality is directly impacted by radiation dose [2], researchers have been actively exploring methods to enhance image quality in low-dose CT scanning. Sparse-view CT scanning emerges as a promising strategy to decrease radiation dose, requiring only a fraction of projection data for image reconstruction [3]. However, the limited number of measurement views leads to reduced acquisition of prior information within the imaged object, consequently resulting in a deterioration in image quality [4]. Therefore, achieving high-quality CT images from sparse-view data is an important area of research.

Sparse-view CT reconstruction typically involves an ill-posed inverse problem. Traditional reconstruction methods like filtered back projection (FBP) [1] yield unsatisfactory results, marked by streaking artifacts and poor image quality. While classical iterative reconstruction algorithms [5,6] improve image quality, the reliance on artificially designed priors that may not match real-world situations comes with the issue of hyperparameter tuning, which in turn affects the generalizability and efficiency of these methods. Deep learning-based reconstruction methods have shown potential in reducing global artifacts and computational costs. Jin *et al*. [7] proposed the FBP-ConvNet algorithm, which incorporates a U-Net network for post-processing in the image domain. Zhang *et al*. [8] developed a DenseNet and deconvolution-based network (DD-Net) that significantly increases network depth while enhancing its expressive power. Chen *et al*. [9] created the residual encoder-decoder convolutional neural network (RED-CNN) by merging autoencoders, deconvolution networks, and shortcut connections, tailored for low-dose CT imaging. Pan *et al*. [10] introduced the multi-domain integrative swin transformer network (MIST-net), utilizing abundant domain features from data, residual data, image, and residual image to enhance the quality of sparse-view CT reconstruction. Despite the impressive reconstruction results, the generality of these supervised methods remains a concern.

Recently, diffusion models, as unsupervised generative models, have gained wide interest in the field of CT reconstruction thanks to their robust capability in mapping complex data distributions [11]. Xia *et al*. [12] proposed a patch-based denoising diffusion probabilistic model (DDPM) for sparse-view CT reconstruction. Xia *et al*. extended this approach to sub-volume-based 3D DDPM [13]. Guan *et al*. [14] introduced a fully score-based generative model in the sinogram domain for sparse-view CT reconstruction. Xu *et al*. [15] presented an innovative approach named the stage-by-stage wavelet optimization refinement diffusion (SWORD) model for sparse-view CT reconstruction. Liu *et al*. [16] proposed an unsupervised sparse-view spectral CT reconstruction and material decomposition algorithm based on the multi-channel score-based generative model (SGM). Du *et al*. [17] introduced the diffusion prior driven neural representation (DPER) that adopted the half quadratic splitting (HQS) algorithm to decompose the inverse problem into data fidelity and distribution prior sub-problems. These diffusion models aim to learn the score function in either the projection or image domains. However, the projection da-

This work was supported by National Natural Science Foundation of China under 621220033 and 62201616. (P. Yu and B. Huang are co-first authors) (Corresponding authors: S. Wang and Q. Liu)
This work did not involve human subjects or animals in its research.

P. Yu and B. Huang are with School of Mathematics and Computer Sciences, Nanchang University, Nanchang 330031, China. ({pengfeiyu, huangbin}@email.ncu.edu.cn)
W. Wu is with the School of Biomedical Engineering, Sun Yat-Sen University, Shenzhen, Guangdong, China. (wuweiw7@mail.sysu.edu.cn)
S. Wang, M. Zhang and Q. Liu are with School of Information Engineering, Nanchang University, Nanchang 330031, China. (shaoyuwang22@gamil.com, {zhangminghui, liuqiegen}@ncu.edu.cn)

taset is expansive and redundant, which can diminish the learning efficiency when the model is applied to raw data, often resulting in images with missing fine details. Additionally, clinical images are intricate, with a variety of scales and details that challenge the capture of their full range of features. The variability in data distribution across different imaging sites and body positions further complicates the task of creating a single, accurate generative model. These issues can cause instability in the diffusion models, affecting the quality of the reconstructed images.

In this study, we present an innovative approach named the ordered-subsets multi-diffusion model (OSMM) for sparse-view CT reconstruction. Specifically, the projection views are separated into several subsets, integrating multi-subsets diffusion model and one-whole diffusion mode. The key idea of this method is to train multiple diffusion models based on the number of divided subsets. The method includes two consecutive stages: multi-subsets diffusion model (MSDM) and one-whole diffusion model (OWDM). At the first stage, multiple diffusion models based on the number of divided subsets are trained. Each diffusion model reconstructs its own sparse subsets of sinogram data, resulting in richer details, providing ample texture and clear details. At the second stage, one-whole diffusion model trained on the entire sinogram is employed to refine finer date features and correct wrong details introduced by MSDM. This two-stage approach enables the model to learn and leverage both whole sinogram and detailed subsets information effectively, contributing to enhance reconstruction quality and accuracy. In summary, the main contributions of this work can be summarized as follows:

● **Enhanced Detail Reconstruction through Subsets Learning:** The OSMM enhances detail reconstruction by dividing the projection data into subsets and utilizing multiple diffusion models for each subset. This approach reduces the complexity of the learning task and improves the reconstruction of fine details within each subset, leading to a more precise reconstruction process in the MSDM.

● **Global Information as a Constraint for Consistent Reconstruction:** By incorporating OWDM with the complete sinogram, the method uses global information as a constraint during the reconstruction process. This constraint helps ensure that the detailed reconstructions from MSDM remain consistent with the global data, thus mitigating the risk of overly random or inconsistent outputs. This approach significantly enhances the overall image quality in sparse-view CT reconstruction by maintaining coherence and structural consistency throughout the generated sinograms.

● **Unsupervised Model Robustness and Generalizability:** The OSMM employs an unsupervised learning approach, providing strong robustness and generalizability. Unlike end-to-end methods that require paired data, the model only need to be trained using full-view sinogram data and can be adapted effectively to different levels of data sparsity in sparse-view sinograms. This ensures consistent performance across varying conditions and enhances its versatility for different clinical scenarios.

The structure of this work is arranged as follows: Section II provides a brief introduction to the relevant knowledge of CT reconstruction. Section III elaborates on the theory and algorithms of CT sparse projection reconstruction based on ordered-subsets multiple diffusion models. Section IV showcases the experimental results and evaluations under ultra-sparse angle conditions. Section V discusses related issues. Finally, this work is summarized in Section VI.

## II. PRELIMINARY

### A. CT Imaging Model

Reconstructing images from incomplete data is inherently an ill-posed inverse problem, posing significant challenges in obtaining accurate solutions. Sparse-view CT reconstruction, in particular, presents significant difficulties. The linear inverse problem in traditional CT imaging can be described as follows:

$$x = AI, \quad (1)$$

where $x$ represents the complete projection data, and $I$ represents the reconstructed image. $A$ is the imaging system matrix.

The transition from complete projection data to sparse-view data can be interpreted as a linear measurement process. Fig. 1 visually illustrates this process. By applying the Radon transform, the original image is converted into projection data. Furthermore, $diag(\Lambda)$ represents the subsampling mask for the projection data, and $P(\Lambda)$ uses this mask to subsample the sinogram into sparse-view data. Consequently, the sparse-view CT reconstruction problem can be formulated as follows:

$$y = P(\Lambda)AI = P(\Lambda)x, \quad (2)$$

where $y$ represents the sparse-view CT projection data, and $x$ denotes the full-view projection data. When processing sparse-view data, direct application of the FBP algorithm for image reconstruction often results in streak artifacts. To obtain higher quality reconstructed images, the full view sinogram needs to be derived from the sparse-view sinogram, which presents an underdetermined inverse problem. To address this challenge, prior information can be incorporated into the regularization objective function by minimizing the following problem:

$$x^* = \arg\min_x \|P(\Lambda)x - y\|_2^2 + \gamma R(x), \quad (3)$$

The objective function consists of two terms. The first term represents data fidelity, ensuring that the sparse-view sinogram aligns with the estimated measurements obtained through the subsampling mask. The second term is the regularization term, where $\gamma$ controls the balance between data fidelity and regularization.

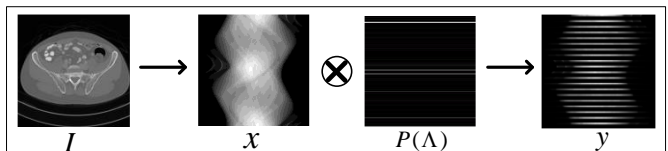

**Fig. 1.** Linear measurement process for sparse-view CT.

### B. Stochastic Differential Equations (SDEs)

Score-based SDE [18] extends the discrete-time framework of denoised diffusion probabilistic models to a continuous-time setting based on SDEs. It also introduces deterministic sampling methods derived from ordinary differential equations.

Song *et al.* [18] were the first to introduce the SDE framework, which includes both the forward diffusion process and its corresponding reverse-time SDE. Wang *et al.* [19] devel-oped an innovative deep generative model based on the Schrödinger bridge concept. To address the challenges posed by Gaussian noise in high-dimensional score-based models, Deasy et al. [20] advanced the denoising score matching technique. Kim et al. [21] introduced a non-linear diffusion process by encoding the image in the latent space and utilizing a trainable normalizing flow model.

Forward SDE: The forward SDE describes a score-based continuous diffusion process through stochastic differential equations. The reverse process is associated with the solution to SDE [22], which includes a drift term for mean shift and a Brownian motion for additional noise:

$$dx = f(x,t)dt + g(t)dw, t \in [0,T], \quad (4)$$

where $w$ is the standard Wiener process, $f(.,t)$ is the drift coefficient, and $g(\cdot)$ is a simplified diffusion coefficient, which is independent of $x$. $p_t(x)$ and $p_T$ denote the marginal and prior distributions respectively. If coefficients are piece-wise continuous, the forward SDE equation has a unique solution [23].

Reversed SDE: Sampling in diffusion models is performed via a corresponding reverse-time SDE of the forward process [24]:

$$dx = [f(x,t) - g(t)^2 \nabla_x \log p_t(x))]d\bar{t} + g(t)d\bar{w},$$
$$t \in [0, T-1], \quad (5)$$

where $\bar{w}$ is again the standard Wiener process running backwards.

Song et al. [18] propose several sampling techniques, including the Predictor-Corrector sampler, which improves sample quality by combining numerical SDE solving with a score-based corrector, such as annealed Langevin dynamics.

### C. Ordered-subsets

Ordered-subsets expectation maximization (OSEM) is an effective and practical reconstruction algorithm, which accelerates convergence by incorporating ordered-subsets [25]. Drawing inspiration from the subset division principle in the OSEM algorithm, we similarly partitioned the projection views into several subsets and independently trained multiple diffusion models on each subset. Notably, we leveraged the subset division concept from the OSEM algorithm for the mining of CT projection data, which not only reduces the difficulty for diffusion models in learning complex CT projection data but also enhances the reconstruction of fine details.

Suppose the total number of projection views is $M$, and we divide them into $N$ subsets, each subset will have $\frac{M}{N}$ views. The subset division is as follows: $x_1 = \{1, N+1, 2\times N+1, 3\times N+1, \cdots, (N-1)\times N+1\}$, $x_2 = \{2, N+2, 2\times N+2, 3\times N+2, \cdots, (N-1)\times N+2\}, \cdots, x_N = \{N, N+N, 2*N+N, 3\times N+N, \cdots, (N-1)\times N+N\}$, The following diagrams illustrate the specific divisions. Fig. 2 shows the case for $N$=2. Fig. 3 shows the case for $N$=3.

The extracting of each subset can be considered as an extraction from the entire projected view with the operator OS. Here, the $n^{\text{th}}$ subset can be expressed as:

$$x_n = \text{OS}(x, n), \quad (6)$$

where $x$ represents the entire projected view and $n$ is the index of the subsets. Subsequently, we obtain $x_1, x_2, \cdots, x_N$, comprising $N$ subsets that allow for independent learning by the diffusion models.

$$\{x_1, x_2, \cdots, x_N\} \subseteq x, \quad (7)$$

and $\text{OS}^-$ is the pseudo-inverse operation of OS. The new subset can be reversely restored to the entire projected view through the $\text{OS}^-$.

$$x = \text{OS}^-(x_1, \cdots, x_N), \quad (8)$$

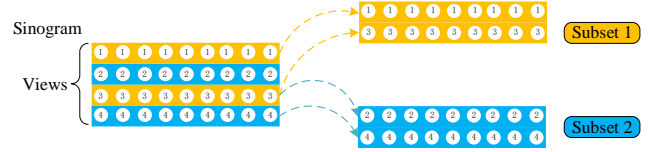

**Fig. 2.** Generation of view-subsets. Sinogram-views are divided among multiple disjoint subsets in a round-robin fashion. In this example, the total views of projection view are split among 2 subsets.

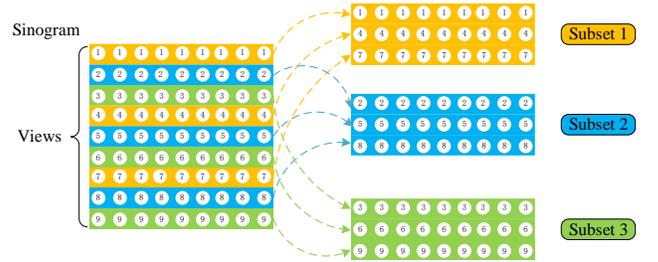

**Fig. 3.** Generation of view-subsets. Sinogram-views are divided among multiple disjoint subsets in a round-robin fashion. In this example, the total views of projection view are split among 3 subsets.

### III. METHODS

#### A. Motivation

Sparse projection data, obtained from limited angles, are more prone to noise compared to complete projection data. While traditional reconstruction algorithms perform well under certain conditions, their performance degrades as the number of projection angles decreases. Sparse-view CT reconstruction often results in streaks and aliasing artifacts, posing significant challenges.

Projection data in CT imaging is crucial as it is the only raw measurement directly related to the physical characteristics of the scanned object. This data preserves fundamental information about the internal structure of the object. However, due to the vast and redundant nature of the entire projection dataset, directly processing this data without preprocessing is not conducive to model learning. Dividing the projection data into multi-subsets can more effectively represent prior information. By learning from these subsets in parallel across multiple diffusion models, this prior information can be better captured and utilized. Such an approach can reduce the difficulty of learning and facilitate the extraction of intrinsic knowledge from the images.

In this work, we propose an innovative method based on ordered-subsets and multiple diffusion models, comprising two consecutive stages: MSDM and OWDM. They extract whole sinogram and detailed subsets information from subset data and complete sinogram respectively. By leveraging their respective strengths, our goal is to approach the ideal image reconstruction effect. Specifically, MSDM extends the sparse projection data, enhancing the reconstruction of fine details while reducing noise and sparse-view artifacts. OWDM, on the other hand, uses global information as a constraint to ensure that the detailed reconstruction from MSDM aligns with the global data, correcting the detail inaccuracies introduced by MSDM and mitigating the risk of randomness or inconsistency. This approach achieves excellent image quality and resolution.

By integrating multiple diffusion models from both MSDM and OWDM, we can effectively learn and utilize both whole sinogram and detailed subsets information, highlight the intrinsic connections between structures and showcase unique advantages in structures, edges, and textures, ultimately improving reconstruction quality and accuracy. As illustrated in Fig. 4, sparse projection data are sequentially reconstructed from left to right to approach the ground truth (GT).

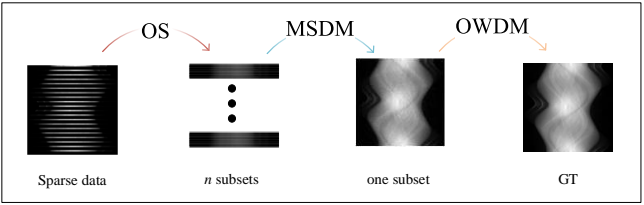

**Fig. 4.** The proposed reconstruction scheme integrates whole sinogram and detailed subsets information. MSDM represents target reconstruction for multi-subsets. OWDM is a global information reconstruction of various image features of a single subset. GT stands for ideal projection data.

### B. Multi-subsets Prior

Considering that multiple models can learn different parts of the data, the representation ability of prior information is enhanced. We propose an innovative approach based on ordered-subsets and multiple diffusion models, rather than relying on a single diffusion model. This approach involves two consecutive training stages: MSDM and OWDM, as shown in Fig. 5. In the MSDM, multiple diffusion models are trained, each focusing on reconstructing its own sparse subset of sinogram data. This results in richer details, providing ample texture and clarity. In the OWDM stage, a one-whole diffusion model is trained to optimize global information, improving image quality and resolution. This two-stage approach allows the model to effectively learn and utilize both whole sinogram and detailed subsets information, thereby enhancing reconstruction quality and accuracy. The projected view is divided into multi-subsets, which can be expressed by the following formula:

$$OS(x, \{1, \cdots, N\}) = x_1, x_2, \cdots, x_N, \quad (9)$$

where OS represents ordered-subsets partitioning. $x_1, x_2, \cdots, x_N$ denote the multi-subsets.

We utilize score-based models with SDEs to train multiple diffusion models in two stages. Score-based models involve both forward SDE and reverse SDE. During the training process, noise is continuously added, and through the encoder and decoder, it gradually transitions from a rough prior distribution to an accurate posterior distribution, ultimately generating a clear reconstructed image.

Consider a continuous diffusion process $\{x(t)\}_{t=0}^{T-1}$ where $x(t) \in \mathbb{R}^n$, with time index $t \in [0, T-1]$ indicating the progression and $n$ representing the image dimension. In this context, the forward SDE process can be expressed as:

$$dx = f(x,t)dt + g(t)dw, \quad (10)$$

where $f(x,t) \in \mathbb{R}^n$ signifies the drift coefficient, and $g(t)$ represents the diffusion coefficient. The term $w \in \mathbb{R}^n$ introduces Brownian motion.

To be specific, we embrace the variance exploding (VE) SDE configuration ($f = 0$, $g = \sqrt{\frac{d[\sigma^2(t)]}{dt}}$), a choice that imparts heightened generative prowess. Incorporating $x_1\ x_2 \cdots x_N$ and $x$ into Eq. (10), the equation metamorphoses into the following forms:

$$dx_1 = \sqrt{d[\sigma^2(t)]/dt}\ dw, \quad (11)$$
$$\vdots$$
$$dx_N = \sqrt{d[\sigma^2(t)]/dt}\ dw, \quad (12)$$
$$dx = \sqrt{d[\sigma^2(t)]/dt}\ dw, \quad (13)$$

where the $\sigma(t) > 0$ denotes a monotonically increasing function, thoughtfully chosen to manifest as a geometric series.

Suppose $x_1(0)$ follows the distribution $p_0$, representing the initial data distribution, while $x_1(T)$ follows $p_T$, symbolizing the acquired prior distribution. Throughout the training phase, the efficacy of the neural network is finely tuned by varying the parameter $\theta_1^*$. This can be viewed as the core objective function within the context of the score-based SDE:

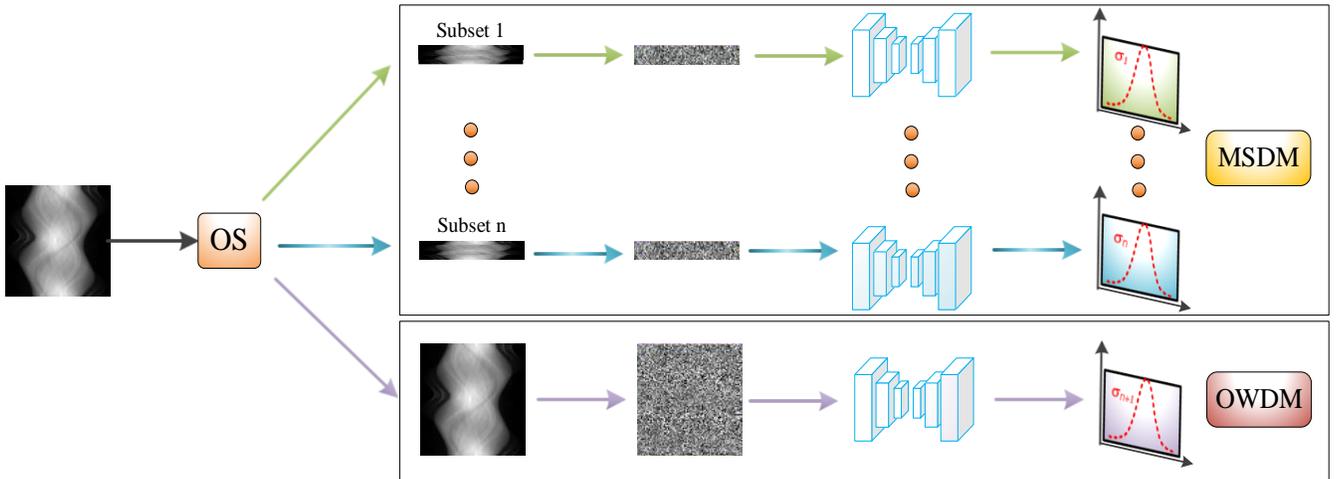

Fig. 5. Visual representation of the forward diffusion procedures of the proposed OSMM. From top to bottom: Elaborate training process for MSDM. Elaborate training process for OWDM.

$$\theta_1^* = \underset{\theta_1}{argmin}\mathbb{E}_t\{\lambda(t)\mathbb{E}_{x_1(0)}\mathbb{E}_{x_1(t)|x_1(0)}[\|s_{\theta_1}(x_1(t),t) -$$
$$\nabla_{x_1(t)} log\, p_t\left(x_1(t)|x_1(0)\right)\|_2^2]\}, \quad (14)$$

where $\lambda(t)$ represents a positively weighted function, and $t$ is uniformly sampled from the interval $[0, T-1]$. The term $p_t(x_1(t)|x_1(0))$ signifies the Gaussian perturbation kernel centered around $x_1(0)$. A neural network $s_{\theta_1}$ is systematically trained to approximate the score, allowing it to capture the essence of $\nabla_{x_1} log\, p_t(x_1)$, i.e., $s_{\theta_1} = \nabla_{x_1} log\, p_t(x_1)$. Once the model is fully trained, the value of $\nabla_{x_1} log\, p_t(x_1)$ can be effectively determined across all time points $t$ through the solution of $s_{\theta_1}(x_1(t), t)$. This provides the model with an ample reservoir of prior knowledge, enabling it to effectively perform image reconstruction. Similarly, the training of $x_2 \cdots x_N$ and $x$ can follow the same procedure to obtain $s_{\theta_2} \cdots s_N$ and $s_{\theta}$.

### C. Iterative Reconstruction

The overall optimization objective of OSMM is minimized as follows:
$$x = \underset{x}{argmin}[\|P(\Lambda)x - y\|_2^2 + r_1R_1(x) + r_2R_2(x)], \quad (15)$$

The first term is data fidelity. To ensure that the output is consistent with the original data, the reconstruction result is processed for data consistency (DC) after each iteration. $r_1$ and $r_2$ are regularization factors for MSDM and OWDM, which are used to balance the data fidelity term and the regularization terms. It obtains the optimal solution by minimizing each element separately. This optimization effort unfolds in multiple consecutive stages, each with a unique contribution to the refinement process. This section describes the iterative reconstruction process of OSMM. Fig. 6 intuitively shows a two-stage iterative method combining MSDM and OWDM for image reconstruction.

**MSDM:** The regularization of MSDM is expressed as follows:
$$R_1(x) = \Sigma_{n=1}^N \lambda_n \|x_n - OS(x, n)\|^2, \quad (16)$$

where $OS(x, n)$ represents the $n^{th}$ subset, $\lambda_n$ are regularization factors for the $n^{th}$ subset. By using multi-subsets for diffusion, each subset gradually reconstructs the image, reducing learning complexity and enhancing detail reconstruction. The overall optimization objective is decomposed into multiple sub-problems, and the subsequent minimization expression of the objective function is as follows:
$$\begin{cases} \|x_1 - OS(x, 1)\|^2, \\ \vdots \\ \|x_N - OS(x, N)\|^2, \end{cases} \quad (17)$$

In the MSDM reconstruction phase, the complete sinogram is divided into N subsets by OS before each iteration. Each subset is then processed by the multiple diffusion model networks, gradually approaching the ideal reconstruction effect. In this way, the high learning difficulty problem can be decomposed into multiple sub-problems, thereby reducing the learning difficulty and enhancing the reconstruction of fine details. The framework for the MSDM iterative reconstruction process is described as follows:
$$\begin{cases} x_1^{t+1}, \cdots, x_N^{t+1} = OS(x^{t+1}, \{1, \cdots, N\}), \\ x_1^t = MSDM_1(x_1^{t+1}), \\ \vdots \\ x_N^t = MSDM_N(x_N^{t+1}), \end{cases} \quad (18)$$

where $MSDM_1 \cdots MSDM_N$ represents multiple networks trained on multi-subsets of different data feature distributions. The equation is an inverse SDE solution problem combined with diffusion priors. Upon the training of the MSDM, obtaining $\nabla_{x_1} log\, p_t(x_1)$ allows for the estimation of the prior distribution of the projection data. The MSDM in the equation is an inverse SDE solution problem combined with diffusion priors. The Predictor (P) solver is used in the model. The Predictor is responsible for sampling from the instantaneous data distribution through a numerical SDE. The discretization of the numerical SDE solver can be expressed as:

$$x_1^t \leftarrow x_1^{t+1} + (\sigma_{t+1}^2 - \sigma_t^2)S_{\theta_1}(x_1^{t+1}, \sigma_{t+1}) + \sqrt{\sigma_{t+1}^2 - \sigma_t^2}z$$
$$t = T-1, \dots, 0, \quad (19)$$

where $z \sim \mathcal{N}(0,1)$ refers to a standard normal distribution, $x(0) \sim p_0$, and $\sigma_0 = 0$ is chosen to simplify the notation. The above formulation is repeated for $t = T-1, \cdots, 0$. Moreover, to ensure that the output remains consistent with the original data, DC processing is performed on the reconstruction results after each step of the P sampler.

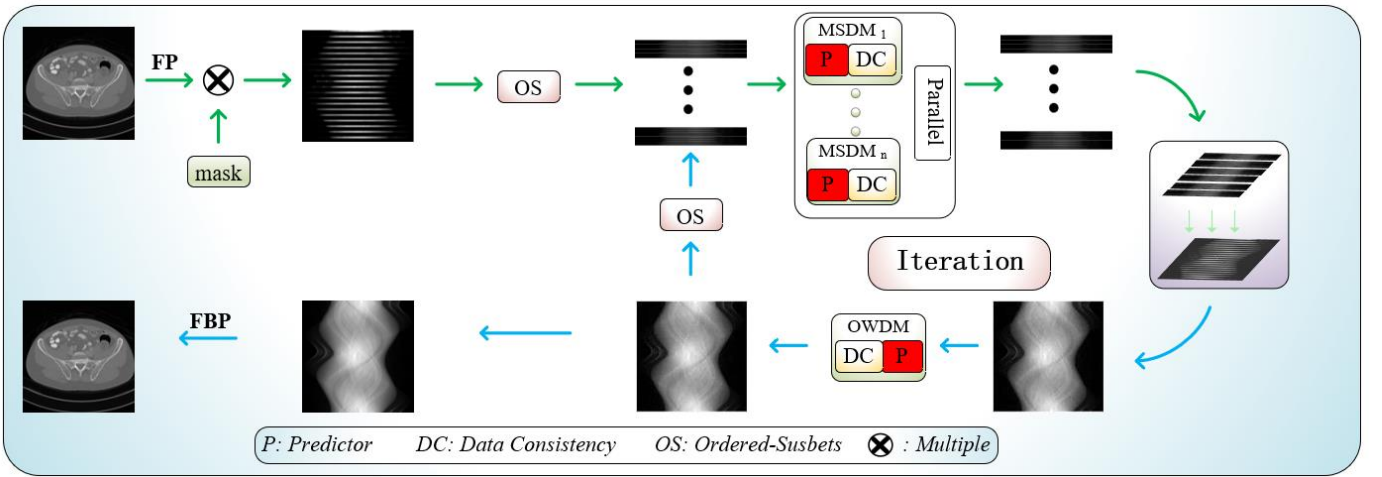

Fig. 6. The reconstruction phase of OSMM in sinogram domain for sparse-view CT. The generation process can be mainly divided into two stages: order-subsets generation, full-view refinement.

*OWDM:* The regularization of OWDM is expressed as follows:

$$R_2(x) = \lambda \|x - OS^-(x_1^t, \cdots, x_N^t)\|^2, \quad (20)$$

where $OS^-(x_1^t, \cdots, x_N^t)$ combines the updated subsets $x_1, x_2, \cdots, x_N$ into a one-whole sinogram. $\lambda$ is regularization factor for the one-whole sinogram.

In the OWDM reconstruction phase, before each iteration, the subsets $x_1, x_2, \cdots, x_N$ reconstructed by MSDM are combined according to the extraction rules or strategies to obtain a whole sinogram. The obtained one-whole sinogram is then input into the OWDM. OWDM also includes a P predictor, followed by DC to ensure that the output is consistent with the original data. The framework description of the OWDM iterative reconstruction process is as follows:

$$\begin{cases} x^t = OS^-(x_1^t, \cdots, x_N^t), \\ x^{t-1} = OWDM(x^t), \end{cases} \quad (21)$$

Finally, the FBP algorithm is used to obtain the reconstructed image. The expression is as follows:

$$\bar{I} = FBP(x), \quad (22)$$

In addition, Algorithm 1 provides detailed operational guidelines for the training and reconstruction phases, ensuring the efficiency and accuracy of the entire reconstruction process.

---

**Algorithm 1: OSMM**

**Training Process**

1: Generating projection $x$;
2: Training datasets construction:
  $x_1, x_2, \cdots, x_N = OS(x, \{1, \cdots, N\})$;
3: Training datasets distribution: $x, x_1, x_2, \cdots x_N$
4: Training with Eq. (14);
5: Output: $s_\theta(x,t), s_{\theta_1}(x_1,t), \cdots, s_{\theta_N}(x_N,t)$

**Inference Process**

**Setting:** $s_\theta, s_{\theta_1}, \cdots, s_{\theta_N}, T, \sigma, \varepsilon$

1: $x_1^T \sim \mathcal{N}(0, \sigma_{max}^2), \cdots, x_N^T \sim \mathcal{N}(0, \sigma_{max}^2)$,
2: For $t = T - 1$ to $0$ **do**
3: For $n = N$ to $1$ **do**
4: Update $x_n^t \leftarrow Predictor(x_n^{t+1}, \sigma_t, \sigma_{t+1})$ with Eq. (18);
5: Update $x_n^t$ by data consistency;
6: $x^t = OS^-(x_1^t, \cdots, x_N^t)$;
7: Update $x^{t-1} \leftarrow Predictor(x^t, \sigma_t, \sigma_{t+1})$ with Eq. (21);
8: Update $x^{t-1}$ by data consistency;
9: End for
10: Return the final reconstructed image $\bar{I}$.

---

## IV. EXPERIMENTS

### A. Data Specification

*AAPM Challenge Data:* This study used simulated abdominal imaging data from the 2016 AAPM CT low-dose grand challenge provided by the Mayo Clinic [26]. The dataset contains high-dose and low-dose CT scans of 10 patients, with 9 patients used for training and 1 patient used for evaluation. We selected 1323 normal-dose CT images with a resolution of $512 \times 512$ and a thickness of 1 mm, dividing them into a training set, and a test set. Specifically, 1186 slices were used for training and 137 were used for testing. The FBP algorithm was a widely recognized neural network architecture interpolating employed to obtain artifact-free images at 720 projection angles, serving as the standard reference. Sparse projection data of 60, 90, 120 and 180 projection angles were extracted from the original 720 projection angle data to evaluate the OSMM. For fan-beam CT reconstructions, sinograms were generated using Siddon's ray-driven algorithm [27, 28]. The distance from the rotation center to the light source and detector is set to 400 and 400. The detector, composed of 720 elements each with a width of 413.

*CIRS Phantom Data:* We obtained a high-quality CT volumetric dataset of size $512 \times 512 \times 100$ voxels, with each voxel of size $0.78 \times 0.78 \times 0.625$, exclusively for testing. The dataset was acquired using a humanoid CIRS phantom on a GE Discovery HD750 CT system with tube current set to 600 mAs. Then 60, 90, 120 and 180 sparse projection view data were extracted from them to evaluate the performance of the algorithm. To further access the robustness and generalization of the proposed method, we utilized the prior knowledge obtained from the AAPM Challenge data to evaluate its performance on the CIRS Phantom dataset.

*Preclinical Mouse Data:* The dead mouse was scanned using a live CT system equipped with a micro-focus X-ray source and a flat-panel X-ray detector. The distances from the source to the detector and object are 1150 $mm$ and 950 $mm$, respectively. The detector consists of 1024×1024 pixels. The size of detector bin is 0.2×0.2 $mm^2$. The projection includes 500 views within the angular range of $[0, 2\pi]$. The reconstructed image is a matrix of 512×512 with 0.15×0.15 $mm^2$ per pixel. Since there are 500 views distributed within the angular range $[0, 2\pi]$, a subsampling factor is set to 10 and 5, and we obtain 50 and 100 views over the scanning range.

### B. Model Training and Parameter Selection

In the experiments, the MSDM and OWDM are trained using the Adam optimization algorithm, as the training process of VE-SDE proposed by Song *et al*. [18]. We initialize the weights using the Kaiming initialization method. The implementation code is programmed in Python, employing the operator discretization library (ODL) [29]. and PyTorch. The computations are carried out on a workstation equipped with an NVIDIA GTX 3090-12GB GPU.

In the proposed multi-scale model architecture, MSDM trains two subset diffusion models, each processing a finite number of views from its own subset of sparse sinogram data. This approach captures intricate details and structures. OWDM trains a diffusion model on the whole sinograms, focusing on learning global information and improving data consistency in the sinogram. The source code is publicly accessible at: https://github.com/yqx7150/OSMM.

### C. Reconstruction Results

In this section, we conduct experiments on the sparse-view CT reconstruction task to demonstrate the superiority of our method. We evaluate the reconstruction performance of five comparative methods against our OSMM on sparse-view CT. The methods compared are FBP [1], U-Net [30], FBPConvNet [7], Patch-based DDPM [12] and GMSD [14]. For the sparse-view CT reconstruction, we utilize 60, 90, 120, and 180 views. FBP is a classic analytical reconstruction algorithm. U-Net is sparsely sampled sinogram to reconstruct CT images by an

FBP algorithm. FBPConvNet combines FBP with deep learning techniques to enhance CT reconstruction, while Patch-based DDPM is built on a diffusion model. GMSD is an advanced generative model. We perform these experiments on three different datasets. The AAPM Challenge data was used for training, while the CIRS Phantom and preclinical mouse datasets were used for generalization tests. Notably, OSMM achieved the best results across all three datasets.

*AAPM Reconstruction Results:* The reconstruction outcomes from the AAPM Challenge dataset were thoroughly assessed, with the average PSNR, SSIM, and MSE values presented in Table I. Compared to the other methods, the average PSNR of the OSMM is the best, as shown in Table I.

The CT images reconstructed from 60 views using the FBP method exhibit poorer quality, blurring crucial details and structures (Fig. 7(b)). Although the U-Net method mitigates some artifacts, it results in the loss of important details (Fig. 7(c)). Additionally, FBPConvNet tends to over-smooth the images, leading to the erasure of many complex details. In contrast, the patch-based DDPM achieves relatively better noise suppression and information preservation (Fig. 7(e)), but still exhibits some loss of detail. The GMSD method performs better overall, but some artifacts are visible in the lower right corner of the area highlighted by the red box (Fig. 7(f)). On the other hand, the images generated by OSMM show superior detail preservation. In the region of interest (ROI), the generated images maintain intricate details without introducing additional artifacts or noise (Fig. 7(g)). Analysis of 90 views reconstructions (Fig. 8) reveals a similar trend to that observed with 90 views. Overall, the results show that OSMM excels in both quantitative evaluations and visual quality, surpassing other methods

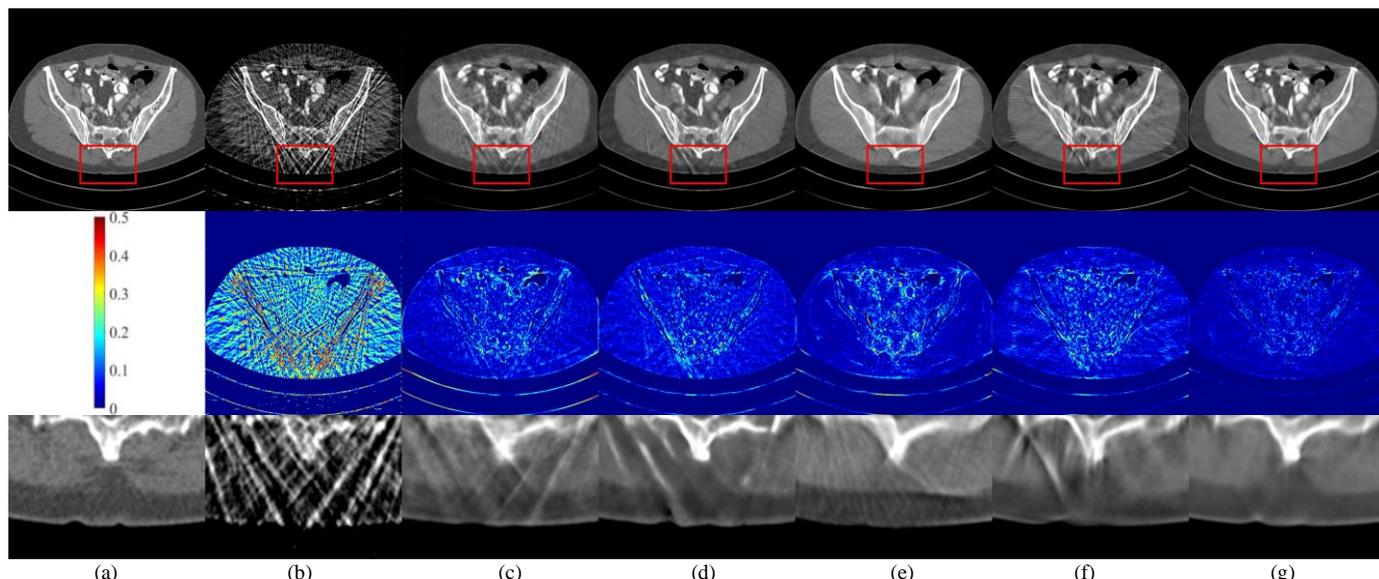

**Fig. 7.** Reconstruction images for AAPM Challenge Data from 60 views using different methods. (a) The reference image versus the images reconstructed by (b) FBP, (c) U-Net, (d) FBPConvNet, (e) Patch-based DDPM, (f) GMSD, and (g)OSMM. Display windows are all set to [-200, 550] HU. The second row shows the difference images, and the third row displays the extracted ROI.

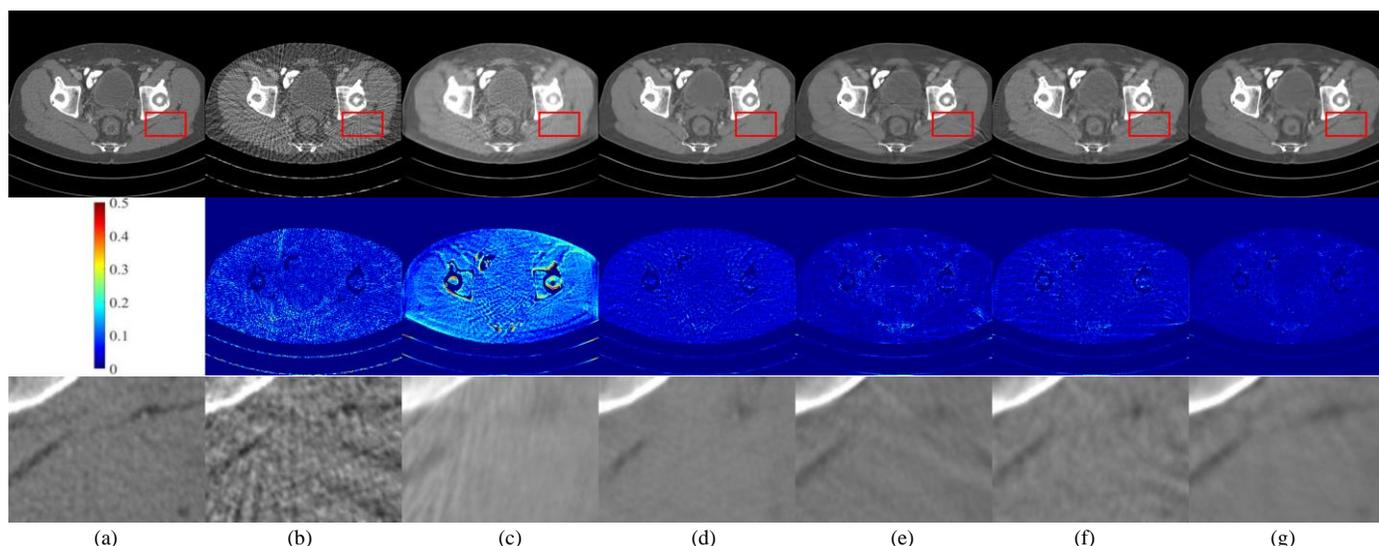

**Fig. 8.** Reconstruction images for AAPM Challenge Data from 90 views using different methods. (a) The reference image versus the images reconstructed by (b) FBP, (c) U-Net, (d) FBPConvNet, (e) Patch-based DDPM, (f) GMSD, and (g)OSMM. Display windows are all set to [-200, 550] HU. The second row shows the difference images, and the third row displays the extracted ROI.

TABLE I
RECONSTRUCTION PSNR/SSIM/MSE OF AAPM CHALLENGE DATA USING DIFFERENT METHODS AT 60, 90, 120, AND 180 VIEWS.

| Views | 60 | 90 | 120 | 180 |
|---|---|---|---|---|
| FBP | 23.18/0.5950/4.88e-3 | 26.20/0.7013/2.45e-3 | 28.30/0.7865/1.52e-3 | 31.69/0.8820/0.70e-3 |
| U-Net | 28.83/0.9365/1.56e-3 | 30.09/0.9472/1.17e-3 | 35.58/0.9765/0.34e-3 | 38.37/0.9853/0.19e-3 |
| FBPConvNet | 35.63/0.9659/0.28e-3 | 37.11/0.9758/0.25e-3 | 39.45/0.9828/0.15e-3 | 42.23/0.9881/0.07e-3 |
| Patch-based DDPM | 32.04/0.9336/0.68e-3 | 35.15/0.9634/0.35e-3 | 37.90/0.9759/0.17e-3 | 40.95/0.9845/0.09e-3 |
| GMSD | 34.31/0.9580/0.41e-3 | 37.25/0.9739/0.20e-3 | 39.41/0.9812/0.12e-3 | 41.44/0.9876/0.08e-3 |
| OSMM | **37.19/0.9749/0.22e-3** | **40.18/0.9835/0.10e-3** | **41.46/0.9869/0.08e-3** | **43.69/0.9917/0.05e-3** |

*CIRS Phantom Reconstruction Results:* To assess the generalization performance of the proposed method, Table II presents the quantitative results derived from the CIRS phantom dataset. Notably, our method outperforms the other techniques, achieving the highest quantitative metrics across all evaluation. Specifically, OSMM exhibits superior structural similarity on the CIRS phantom dataset. Compared to Patch-based DDPM, OSMM achieves improvements of 9.78 dB, 10.01 dB, 8.62 dB, and 7.75 dB at 60, 90, 120, and 180 views, respectively. Similarly, when compared to GMSD, OSMM shows gains of 6.80 dB, 8.84 dB, 8.60 dB, and 4.01 dB at 60, 90, 120, and 180 views, respectively.

Fig. 9 displays the reconstruction results of all algorithms on the CIRS phantom dataset. The FBP method introduces significant streak artifacts, obscuring key image structures. Although U-Net and FBPConvNet reduce streak artifacts and eliminate much of the noise, they still suffer from blurred outlines and unclear details, resulting in less than optimal visual outcomes. The Patch-based DDPM and GMSD methods achieve notable artifact reduction, but this comes at the expense of some texture information. Overall, these substantial improvements across various metrics and evaluation criteria highlight the effectiveness of OSMM, which consistently delivers higher accuracy, better detail preservation, and sharper image reconstructions.

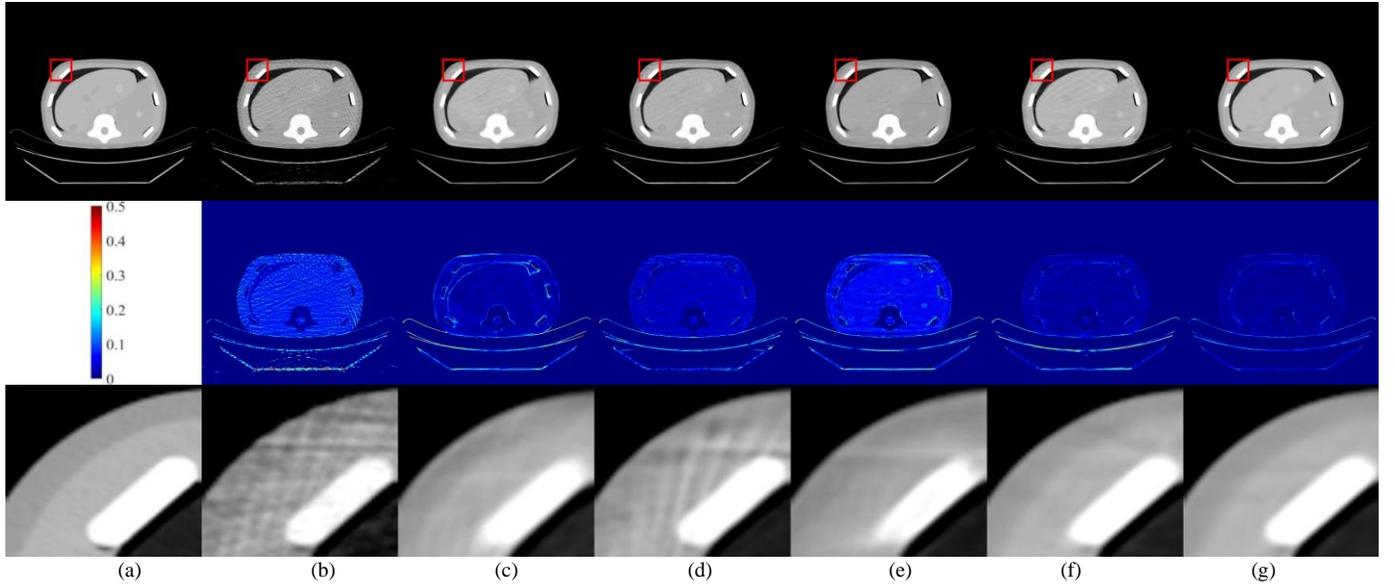

**Fig. 9.** Reconstruction images CIRS Challenge Data from 90 views using different methods. (a) The reference image versus the images reconstructed by (b) FBP, (c) U-Net, (d) FBPConvNet, (e) Patch-based DDPM, (f) GMSD, and (g)OSMM. Display windows are all set to [-200, 550] HU. The second row shows the difference images, and the third row displays the extracted ROI.

TABLE II
RECONSTRUCTION PSNR/SSIM/MSE OF CIRS CHALLENGE DATA USING DIFFERENT METHODS AT 60, 90, 120, AND 180 VIEWS.

| Views | 60 | 90 | 120 | 180 |
|---|---|---|---|---|
| FBP | 17.96/0.5028/16.00e-3 | 23.71/0.5982/4.26e-3 | 25.42/0.6827/2.87 e-3 | 28.83/0.7994/1.31e-3 |
| U-Net | 25.76/0.8883/2.66e-3 | 31.62/0.9542/0.67e-3 | 35.57/0.9727/0.28e-3 | 38.31/0.9852/0.15e-3 |
| FBPConvNet | 26.70/0.9333/2.28e-3 | 32.11/0.9579/0.61e-3 | 35.30/0.9730/0.35e-3 | 38.80/0.9848/0.11e-3 |
| Patch-based DDPM | 25.93/0.8979/2.57e-3 | 31.51/0.9676/0.73e-3 | 34.77/0.9781/0.34e-3 | 39.12/0.9895/0.12e-3 |
| GMSD | 28.19/0.9209/1.53e-3 | 32.86/0.9603/0.52e-3 | 38.00/0.9815/0.16e-3 | 42.86/0.9911/0.05e-3 |
| OSMM | **35.71/0.9826/0.28e-3** | **41.52/0.9920/0.07e-3** | **43.37/0.9940/0.05e-3** | **46.87/0.9966/0.21e-3** |

*Preclinical Mouse Reconstruction Results:* To evaluate the efficiency on real data, the network within OSMM was trained on the AAPM challenge data and tested on preclinical mouse data. Fig. 10 depicts the images reconstructed by different methods. The reference reconstruction was obtained from full-view projections. Fig. 10 clearly highlights the persistent noise present in FBP reconstructions. Inspection of the extracted ROIs shows that both U-Net and FBPConvNet tend to produce blurred edges, compromising the clarity of details within the ROIs. In contrast, patch-based DDPM and GSDM exhibit varying degrees of detail loss in their reconstruction results. Meanwhile, OSMM consistently stands out by preserving sharper edges and maintaining superior accuracy in capturing details compared to other methods. The comparison of reconstructed images highlights the continued excellence of OSMM compared to other methods.

Table III presents a detailed quantitative analysis of the preclinical mouse data, emphasizing the improvements of OSMM over GMSD across various metrics. Specifically, OSMM achieved gains of 3.41 dB and 1.13 dB at 50 and 100 views, respectively. OSMM consistently delivers enhanced edge clarity and superior accuracy in detail representation, reaffirming its superiority over other methods.

TABLE III
RECONSTRUCTION PSNR/SSIM/MSE OF PRECLINICAL MOUSE DATA USING DIFFERENT METHODS AT 50 AND 100 VIEWS.

| Views | 50 | 100 |
|---|---|---|
| FBP | 27.70/0.6269/1.69e-3 | 31.31/0.7631/0.74e-3 |
| U-Net | 28.81/0.9193/1.31e-3 | 34.08/0.9616/0.39e-3 |
| FBPConvNet | 28.98/0.6575/1.26e-3 | 34.46/0.8854/0.36e-3 |
| Patch-based DDPM | 33.98/0.9168/0.40e-3 | 39.89/0.9702/0.10e-3 |
| GMSD | 35.26/0.9379/0.30e-3 | 39.07/0.9714/0.12e-3 |
| OSMM | **38.67/0.9625/0.13e-3** | **40.20/0.9752/0.10e-3** |

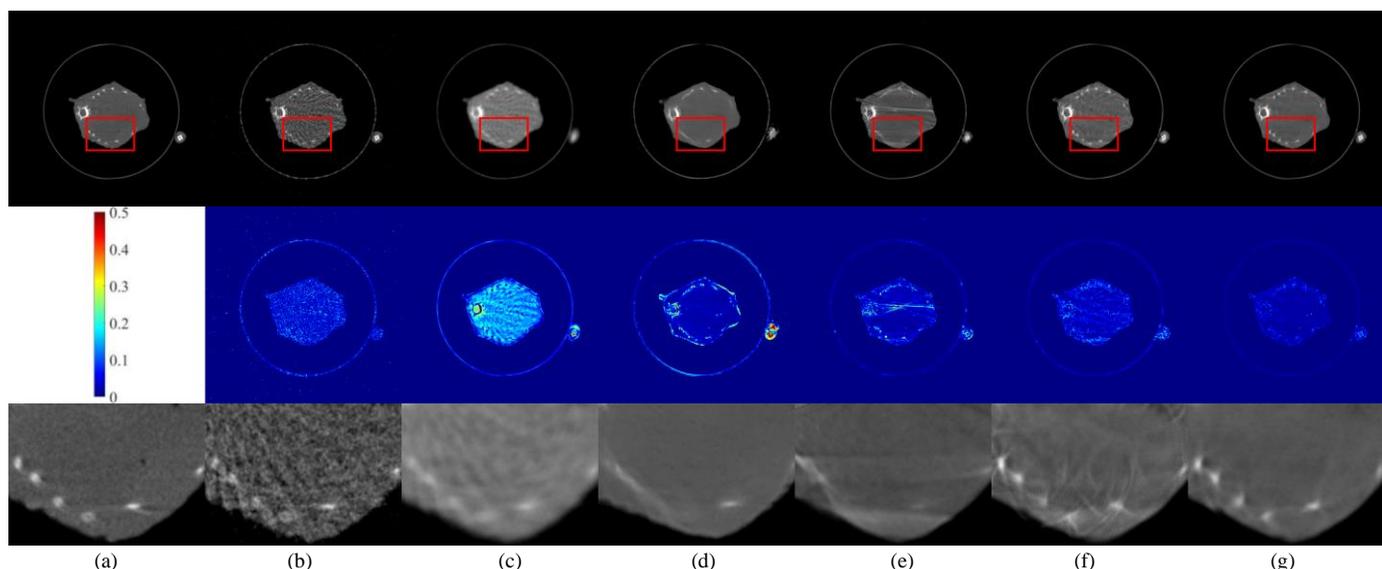

Fig. 10. Reconstruction images from 50 views using different methods. (a) The reference image versus the images reconstructed by (b) FBP, (c) U-Net, (d) FBP-ConvNet, (e) Patch-based DDPM, (f) GMSD, and (g)OSMM. Display windows are all set to [-200, 550] HU. The second row shows the difference images, and the third row displays the extracted ROI.

TABLE IV
RECONSTRUCTION RESULTS ON AAPM CHALLENGE DATA

| Methods | Views | PSNR | SSIM | MSE |
|---|---|---|---|---|
| MSDM | 60 | 35.67 | 0.9651 | 0.30e-3 |
| | 90 | 38.76 | 0.9786 | 0.15e-3 |
| | 120 | 40.16 | 0.9841 | 0.12e-3 |
| | 180 | 43.08 | 0.9903 | 0.05e-3 |
| OWDM | 60 | 36.13 | 0.9698 | 0.29e-3 |
| | 90 | 39.26 | 0.9787 | 0.15e-3 |
| | 120 | 40.16 | 0.9841 | 0.12e-3 |
| | 180 | 43.09 | 0.9903 | 0.05e-3 |
| OSMM | 60 | **37.19** | **0.9749** | **0.22e-3** |
| | 90 | **40.18** | **0.9835** | **0.10e-3** |
| | 120 | **41.46** | **0.9869** | **0.08e-3** |
| | 180 | **43.69** | **0.9917** | **0.05e-3** |

### D. Ablation Study

This work introduces an OSMM that integrates the strengths of both MSDM and OWDM, focusing on capturing the global consistency and precise local details in the reconstructed image. The individual contributions of MSDM and OWDM to the reconstruction results are examined through detailed ablation experiments.

As shown in Table IV, the significant advantage of enhancing the image reconstruction performance by using ordered-subsets multi-diffusion models is clearly demonstrated. The combination of MSDM and OWDM not only preserves the respective benefits of each model but also substantially improves overall reconstruction quality.

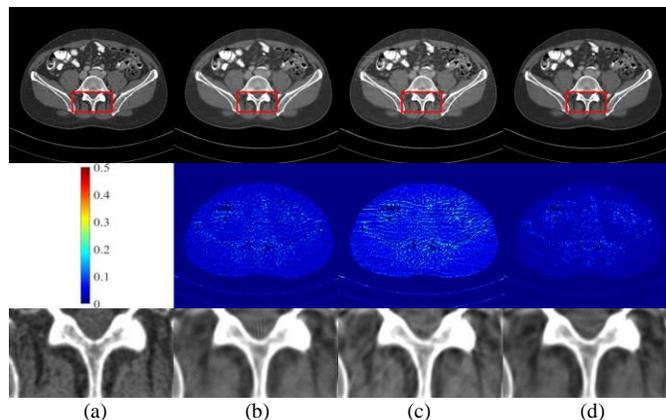

Fig. 11. Reconstructed images from 90 different perspectives obtained using different methods: (a) GT, (b) OSMM without the MSDM, (c) OSMM without the OWDM, and (d) fully implemented OSMM. Display windows are all set to [-200, 550] HU. The second and third rows depict difference images and extracted ROI.

## V. DISCUSSION AND CONCLUSION

In this paper, we propose an ordered-subsets multiple diffusion model for sparse view CT reconstruction. The OSMM innovatively divides CT projection data into equal subsets and employs multiple diffusion models to learn independently from each subset. This targeted learning approach reduces complexity and enhances the reconstruction of fine details.

TABLE V
RECONSTRUCTION RESULTS ON AAPM CHALLENGE DATA

| Number | Views | PSNR | SSIM | MSE |
|---|---|---|---|---|
| 2 | 60 | 37.19 | **0.9749** | 0.22e-3 |
| | 90 | **40.18** | 0.9835 | **0.10e-3** |
| | 120 | **41.46** | 0.9869 | **0.08e-3** |
| | 180 | **43.69** | 0.9917 | **0.05e-3** |
| 3 | 60 | **37.64** | 0.9743 | **0.19e-3** |
| | 90 | 39.78 | **0.9872** | 0.11e-3 |
| | 120 | 41.27 | 0.9869 | 0.08e-3 |
| | 180 | 43.15 | 0.9915 | 0.05e-3 |

To verify the impact of the number of subsets on image reconstruction quality, we divided the projection data into two and three subsets. Table V presents the average test results of models with different numbers of subsets on the AAPM challenge data. Specifically, when projection data is noisy, using too many subsets can hinder convergence and lead to suboptimal reconstruction results [31]. Additionally, increased computational complexity due to multiple diffusion models presents a challenge, which we plan to address in future work.

In conclusion, the OSMM significantly enhances sparse-view CT reconstruction by improving detail preservation, maintaining consistency with global data, and demonstrating robust generalizability. By employing subset learning, the method effectively reconstructs fine details. Incorporating global information as a constraint ensures that the reconstructed details remain coherent and consistent with the overall dataset. Furthermore, the use of an unsupervised learning approach increases the model's adaptability and performance across varying levels of data sparsity, making it a versatile solution for diverse clinical scenarios.

## ACKNOWLEDGMENTS

The authors sincerely thank the anonymous referees for their valuable comments on this work. All authors declare that they have no known conflicts of interest in terms of competing financial interests or personal relationships.